\documentclass[aps,prl,twocolumn,showpacs,citeautoscript,reprint,longbibliography]{revtex4-2}
\usepackage{boldline,multirow}
\usepackage{float}
\usepackage{graphicx,natbib}
\usepackage{bm,amsmath,amssymb,wasysym,amsfonts,dcolumn}
\usepackage[colorlinks=true, urlcolor=blue, linkcolor=blue, citecolor=blue, pdfborder={0 0 0}]{hyperref}
\usepackage{ulem}
\usepackage[usenames,dvipsnames]{xcolor}
\usepackage{lipsum}
\usepackage{epstopdf}
\usepackage{cleveref}
\crefname{figure}{Fig.}{Figs}
\crefname{table}{Table}{Tables}
\crefname{section}{Sec.}{Sections}

\begin{document}

%%%%%%%%10%%%%%%%%20%%%%%%%%30%%%%%%%%40%%%%%%%%50%%%%%%%%60%%%%%%%%70%%%%%%%%80
\title{Spin Hall Effect in a Thin Pt Film}
		
\author{R. S. Nair}
\affiliation{Faculty of Science and Technology and MESA$^+$ Institute for Nanotechnology, University of Twente, P.O. Box 217,
	7500 AE Enschede, The Netherlands}
	
\author{M. S. Rang}
\affiliation{Faculty of Science and Technology and MESA$^+$ Institute for Nanotechnology, University of Twente, P.O. Box 217,
	7500 AE Enschede, The Netherlands}

%\author{\color{red}R.J.H. Wesselink}
%\affiliation{Faculty of Science and Technology and MESA$^+$ Institute for Nanotechnology, University of Twente, P.O. Box 217,
%		7500 AE Enschede, The Netherlands}
%
%\author{\color{red}Z. Yuan}
%\affiliation{\color{red}The Center for Advanced Quantum Studies and Department of Physics, Beijing Normal University, 100875 Beijing, China}	
	
\author{Paul J. Kelly\thanks{corresponding author}}
\email[Email: ]{P.J.Kelly@utwente.nl}
\affiliation{Faculty of Science and Technology and MESA$^+$ Institute for Nanotechnology, University of Twente, P.O. Box 217,
	7500 AE Enschede, The Netherlands}

\date{\today}

\begin{abstract}
A density-functional-theory based relativistic scattering formalism is used to study charge transport through thin Pt films with room temperature lattice disorder. A Fuchs-Sondheimer specularity coefficient $p \sim 0.5$ is needed to describe the suppression of the charge current at the surface even in the absence of surface roughness. 
The charge current drives a spin Hall current perpendicular to the surface. Analysing the latter with a model that is universally used to interpret the spin Hall effect in thin films and layered materials, we are unable to recover values of the spin-flip diffusion length $l_{\rm sf}$ and spin Hall angle $\Theta_{\rm sH}$ that we obtain for bulk Pt using the same approximations. We trace this to the boundary conditions used and develop a generalized  model that takes surface effects into account. A reduced value of $\Theta_{\rm sH}$ at the surface is then found to describe the first-principles transport results extremely well. The in-plane spin Hall effect is substantially enhanced at the surface.
\end{abstract}

\pacs{}

\maketitle

\begin{figure}[b]
\centering
\includegraphics[width=8.6cm]{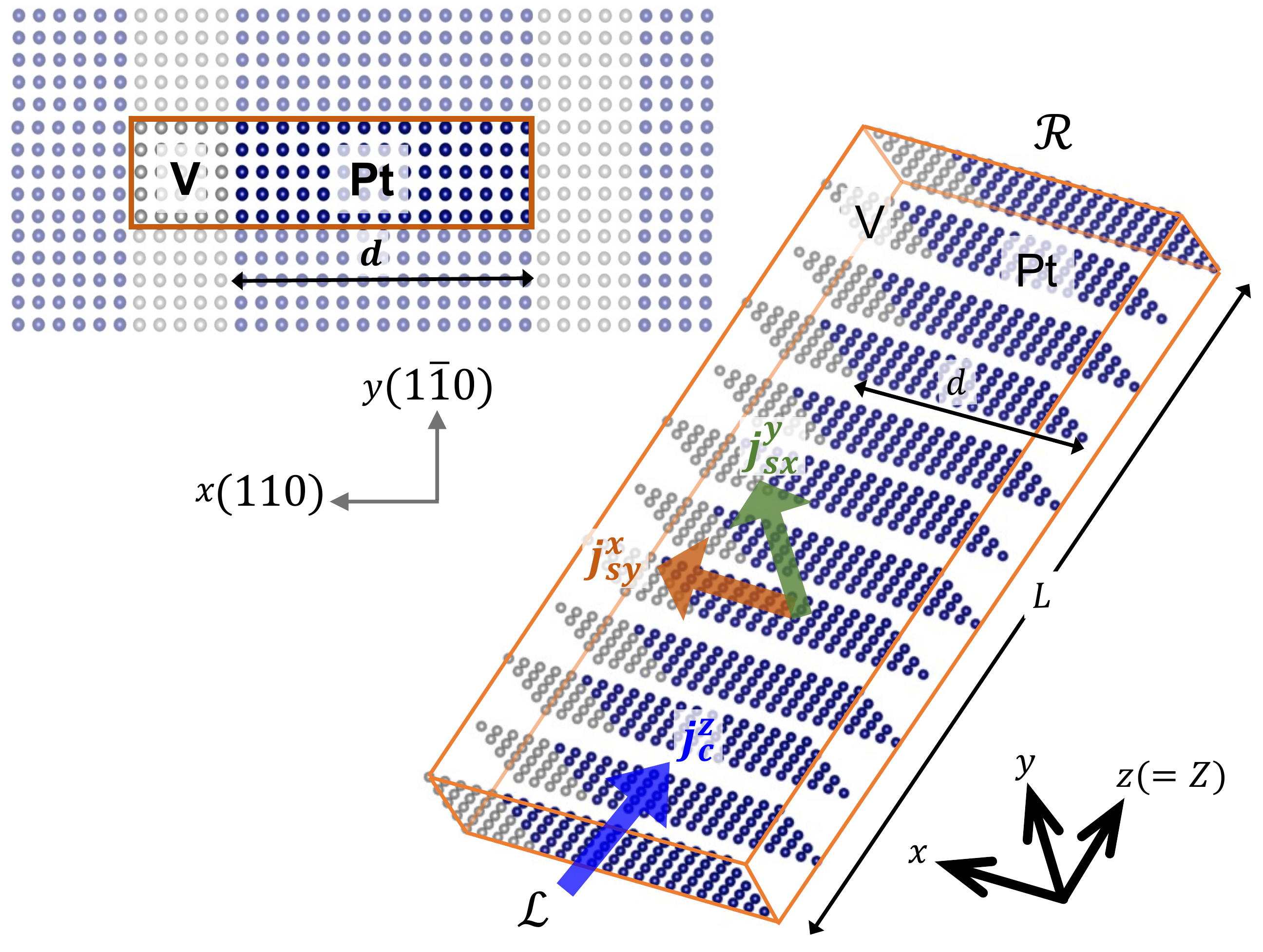}
\caption{Schematic of the scattering geometry used to study transport in a thermally disordered, [110]-oriented thin film of fcc Pt (blue). The thin film of thickness $d$ and length $L$ is sandwiched between semiinfinite Pt leads $\mathcal{L}$ and $\mathcal{R}$ in the $z$ direction. The layer separation in the $z$ direction is exaggerated for clarity. The vacuum region is modelled using a thickness of vacuum equal to five layers of ``empty'' spheres (grey) in the $x$ direction and superlattice periodicity in the $x$ and $y$ directions is assumed. A single [001] plane of the scattering geometry is shown on the top left with the ``lateral supercell'' indicated in orange. A charge current $j^z_c$ is passed between the leads  $\mathcal{L}$ and $\mathcal{R}$ that excites spin currents $j^x_{sy}$ (orange arrow) in the $x$ direction which accumulates at the surface and $j^y_{sx}$ (green arrow) flowing in the $y$ direction parallel to the surface. 
	}
\label{fig:geometry}
\end{figure}

%%%%%%%%%10%%%%%%%%20%%%%%%%%30%%%%%%%%40%%%%%%%%50%%%%%%%%60%%%%%%%%70%%%%%%%%80
{\color{red}\it Introduction.---}The passage of a charge current through a heavy metal excites orthogonal spin currents because of the asymmetric spin scattering mediated by spin-orbit coupling, the well-known spin Hall effect (SHE) \cite{Dyakonov:zetf71, *Dyakonov:pla71, Hirsch:prl99, Hoffmann:ieeem13, Sinova:rmp15}. At an interface or surface which breaks the translational symmetry, spin accumulates on a length scale of the spin-flip diffusion length (SDL) $l_{\rm sf}$ \cite{Zhang:prl00}. 
For a thin film of finite thickness $d$ in the $x$ direction, see \cref{fig:geometry}, the diffusive spin Hall current $j^x_{sy}(x)$ towards the surface at $x=d/2$ that is generated in response to a charge current $j^z_c$ applied in the $z$ direction and the resulting spin accumulation $\mu^x_{sy}(x)$ are described by \cite{Valet:prb93, Zhang:prl00}
\begin{subequations}
\label{eq1}
\begin{align}
j^x_{sy}(x) &=\Theta_{\rm sH}j^z_c 
\bigg[1-\frac{\cosh(x/l_{\rm sf})}{\cosh(d/2l_{\rm sf})} \bigg] , 
\label{eq1a}\\
\mu^x_{sy}(x) & =2\Theta_{\rm sH} j^z_c \rho \, l_{\rm sf}   
\bigg[\frac{\sinh(x/l_{\rm sf})}{\cosh(d/2l_{\rm sf})} \bigg] .
\label{eq1b}
\end{align}
\end{subequations}
Here, the spin current in the $x$ direction is polarized in the $y$ direction and the spin Hall angle (SHA) $\Theta_{\rm sH}$ characterizes the efficiency of charge-to-spin conversion. This model is expected to hold for $d > l_{\rm sf}$ and is used to extract values of $l_{\rm sf}$ and $\Theta_{\rm sH}$ from experiments on NM$|$FM bilayers comprising nonmagnetic (NM) and ferromagnetic (FM) metals using spin-pumping and the inverse spin Hall effect (ISHE) \cite{Saitoh:apl06, Ando:prl08, Mosendz:prl10} or the SHE and spin transfer torque \cite{Liu:prl11}, or using the nonlocal spin injection method \cite{Kimura:prl07, Vila:prl07}. The NM metal that is most frequently studied is Pt whose thickness is typically chosen in the range 10-20~nm. 
The values of $l_{\rm sf}$ and $\Theta_{\rm sH}$ reported for Pt span an order of magnitude \cite{Bass:jpcm07, Sinova:rmp15}. All three experimental methods rely on interfaces to inject or detect the spin current and a failure to take account of interface spin-flipping 
%, as described by the spin memory loss (SML) parameter $\delta$, 
was suggested \cite{Rojas-Sanchez:prl14} as the reason for the wide spread. However, taking it into account has not brought consensus any closer about the values of the SDL and SHA \cite{Wesselink:prb19}. 

Recently the magneto-optical Kerr effect (MOKE) was used to probe the spin accumulation in a thin film and thereby estimate $\Theta_{\rm sH}$  and $l_{\rm sf}$ without introducing an interface between Pt and another metal \cite{Stamm:prl17}. However, the values of $\Theta_{\rm sH}$ and $l_{\rm sf}$ extracted from these experiments for Pt are twice those of the best available theoretical estimates using ab-initio calculations \cite{Wesselink:prb19, Nair:prl21} motivating us to examine the SHE in thin films in more detail. 

%%%%%%%%%10%%%%%%%%20%%%%%%%%30%%%%%%%%40%%%%%%%%50%%%%%%%%60%%%%%%%%70%%%%%%%%80
{\color{red}\it Method.---}A fully relativistic quantum mechanical scattering scheme was used to study transport in the thermally disordered, [110]-oriented, thin Pt film sketched in \cref{fig:geometry}. The disordered Pt film is  sandwiched between left ($\mathcal{L}$) and right ($\mathcal{R}$) Pt leads in the $+z$ and $-z$ directions. In the first step of a two-stage procedure, Bloch eigenstates at the Fermi energy of the crystalline Pt leads are determined and classified according to whether they propagate in the $+z$ or $-z$ direction. In the second step, these scattering states are used as boundary conditions to find quantum mechanical solutions throughout the scattering region using Ando's wave-function matching scheme \cite{Ando:prb91} implemented \cite{Starikov:prl10, *Starikov:prb18} using a very efficient basis of tight-binding muffin tin orbitals \cite{Andersen:prl84, *Andersen:85, *Andersen:prb86}. 
Periodic boundary conditions are used in the $x$ and $y$ directions whereby use can be made of Bloch's theorem to efficiently determine and keep track of the lead states \cite{Xia:prb06}. The resulting periodic images of the thin film are separated by vacuum that is represented in the atomic spheres approximation (ASA) using ``empty spheres'' \cite{Skriver:prb92a, *Skriver:prb92b, Daalderop:prb94}. For the density functional theory calculations, we used the local spin density approximation as parameterised by von Barth and Hedin (vBH) \cite{vonBarth:jpc72}, the experimental lattice constant, an $spd$ basis and omit three-centre terms in the SOC matrix. The potential calculation was iterated to self consistency with SOC included \cite{footnote4, footnote5}.

A charge current $j_c^z(x)$ is passed through the thin diffusive Pt film in the $+z$ direction by applying an infinitesimal voltage difference between the Pt leads so that at the Fermi energy only Bloch states propagating in the $+z$ direction are occupied in the $\mathcal{L}$ lead and Bloch states propagating in the $-z$ direction are unoccupied in the $\mathcal{R}$ lead.  From the results of the scattering calculations, spin currents in the $x$ and $y$ directions are calculated as in Ref.~\cite{Wesselink:prb19}; their interpolation onto a three dimensional grid is described in Ref.~\cite{Nair:prb21a}.

%%%%%%%%%10%%%%%%%%20%%%%%%%%30%%%%%%%%40%%%%%%%%50%%%%%%%%60%%%%%%%%70%%%%%%%%80

\begin{figure}[!t]
\includegraphics[width=8.6cm]{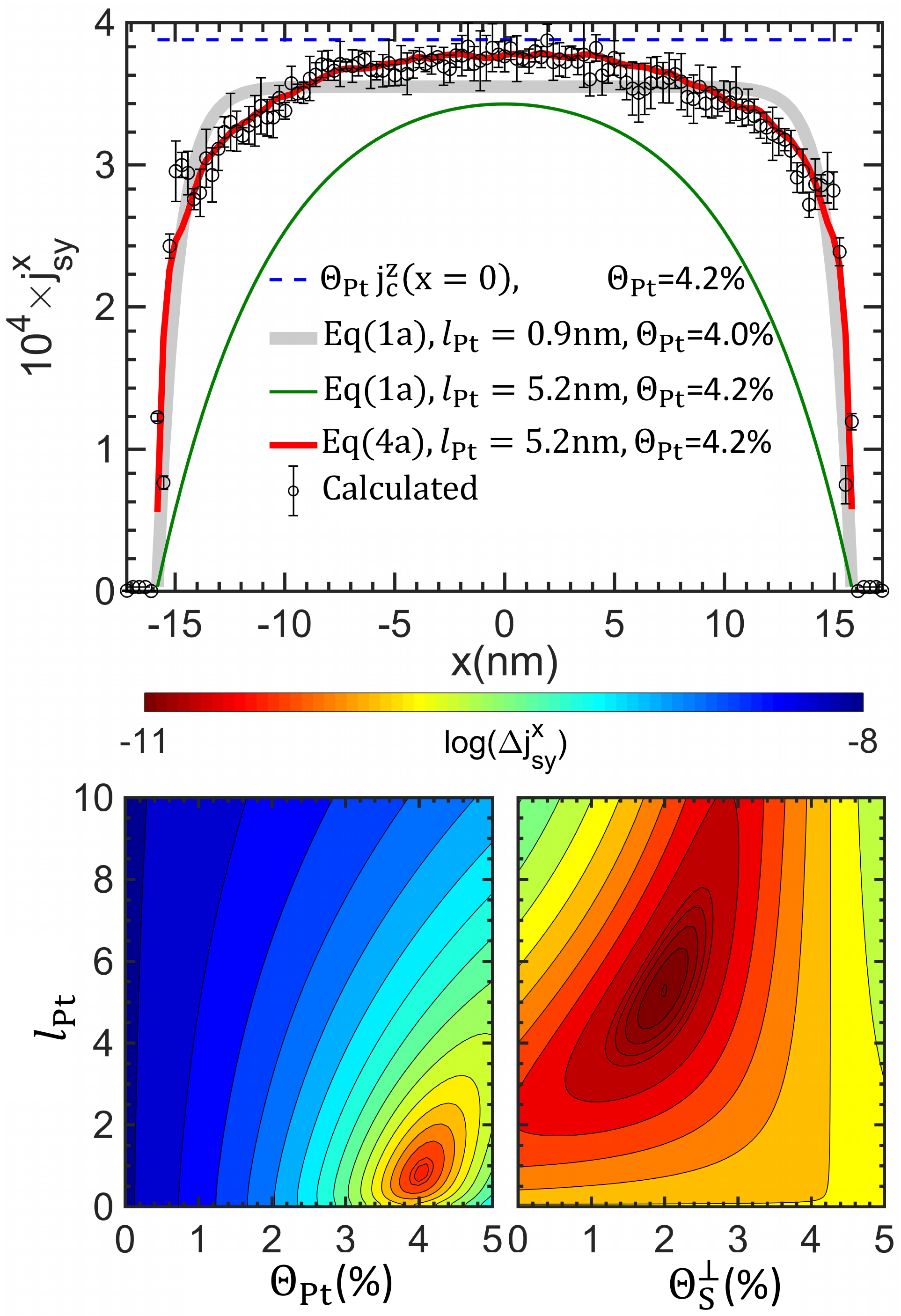}
%\llap{\parbox[b]{5.25in}{\Large(a)\\\rule{0ex}{4.30in}}}\llap{\parbox[b]{5.25in}{\Large(b)\\\rule{0ex}{2.15in}}}
\caption{(Top) Spin current $j^x_{sy}$ flowing in the $x$ direction at room temperature as a function of $x$ (circles). The green curve results from \eqref{eq1a} using the ``bulk'' values $l_{\rm Pt}=5.2\,$nm and {\color{blue}$\Theta_{\rm Pt}=4.2\%$} calculated for bulk Pt \cite{Wesselink:prb19, Nair:prl21}. 
The grey curve is the best fit using \eqref{eq1a} with $l_{\rm Pt}=0.9\,$nm and $\Theta_{\rm Pt}=4.0\%$.
The red curve is obtained by taking the calculated $j_c^z(x)$, bulk values of $l_{\rm Pt}=5.2\,$nm and $\Theta_{\rm Pt}=4.2\%$ and fitting with \eqref{eq4a} and \eqref{eq5} to obtain $\Theta_{\rm S}=2.54\pm0.03\%$.
The horizontal dashed blue line represents the bulk value $\Theta_{\rm Pt}=4.2\%$ for the charge current density at the centre of the thin film. 
Error bars on the calculated data represent the average deviation over five configurations of thermal disorder. 
(Bottom) Logarithm of the root mean square residuals using (left) \eqref{eq1a} to describe the ab-initio calculated data in the $\{l_{\rm Pt},\Theta_{\rm Pt}\}$ parameter space and (right) \eqref{eq4a} and \eqref{eq5} in the $\{l_{\rm Pt},\Theta_{\rm S}\}$ parameter space with the bulk value of $\Theta_{\rm Pt}=4.2 \%$.
}
\label{fig2}
\end{figure}

{\color{red}\it Calculations.---}The results we find for the spin Hall current $j^x_{sy}(x)$ flowing towards the surface of the thin film sketched in \cref{fig:geometry} are presented in \cref{fig2} for a free standing Pt slab of thickness $d=31.60 \,$nm corresponding to 115 atomic layers in the $x$ direction that are separated from their periodic images by five layers of empty spheres. 90 layers of this 360=(115+5)$\times 3$ supercell were stacked in the $z$ direction and the atoms in the scattering region were displaced at random with a Gaussian distribution of atomic displacements chosen to reproduce the experimental bulk resistivity of $\rho=10.8 \, \mu \Omega \,$cm \cite{LiuY:prb11, *LiuY:prb15, HCP90} at room temperature (RT=$\rm 300K$). The thickness of the slab is $d \approx 6 \, l_{\rm Pt}$ for a value of $l_{\rm Pt} \equiv l_{\rm sf}^{\rm Pt} \approx 5.2 \,$nm estimated from the decay of a fully polarized spin current injected into Pt \cite{Wesselink:prb19, Nair:prl21} making it essentially bulk-like. Consistent with this is that at the center of the slab, the spin Hall angle (the ratio of the transverse spin current measured in units of $\hbar/2$ to the longitudinal charge current measured in units of the electron charge $-e$) is the bulk value, $\Theta_{\rm Pt} \equiv \Theta_{\rm sH}^{\rm Pt} \approx 4.2 \%$ \cite{Wesselink:prb19}. Small oscillations that are discernible in the calculated profiles are attributable to standing waves that are not described by semi-classical (Boltzmann) descriptions of transport. Using the bulk values of $l_{\rm Pt}$ and $\Theta_{\rm Pt}$ in \eqref{eq1a} yields the green curve in \cref{fig2} that is clearly a poor representation of the ab-initio results. If instead we perform a least squares fit in the $\{l_{\rm Pt},\Theta_{\rm Pt}\}$ parameter space (bottom left panel), we find a best fit with $l_{\rm Pt}=0.9\,$nm and $\Theta_{\rm Pt}=4.0 \%$ yielding the grey curve in \cref{fig2} (top panel). Although we know what $l_{\rm Pt}$ and $\Theta_{\rm Pt}$ should be from our bulk calculations \cite{Wesselink:prb19, Nair:prl21}, we are unable to recover these values from the present calculations for a bulk-like thin film carried out with the same approximations. We identify the boundary condition of \eqref{eq1a} and \eqref{eq1b} used to extract $l_{\rm Pt}$ and $\Theta_{\rm Pt}$ from the spin current as the culprit.

\begin{figure}[t]
\centering
\includegraphics[width=8.6cm]{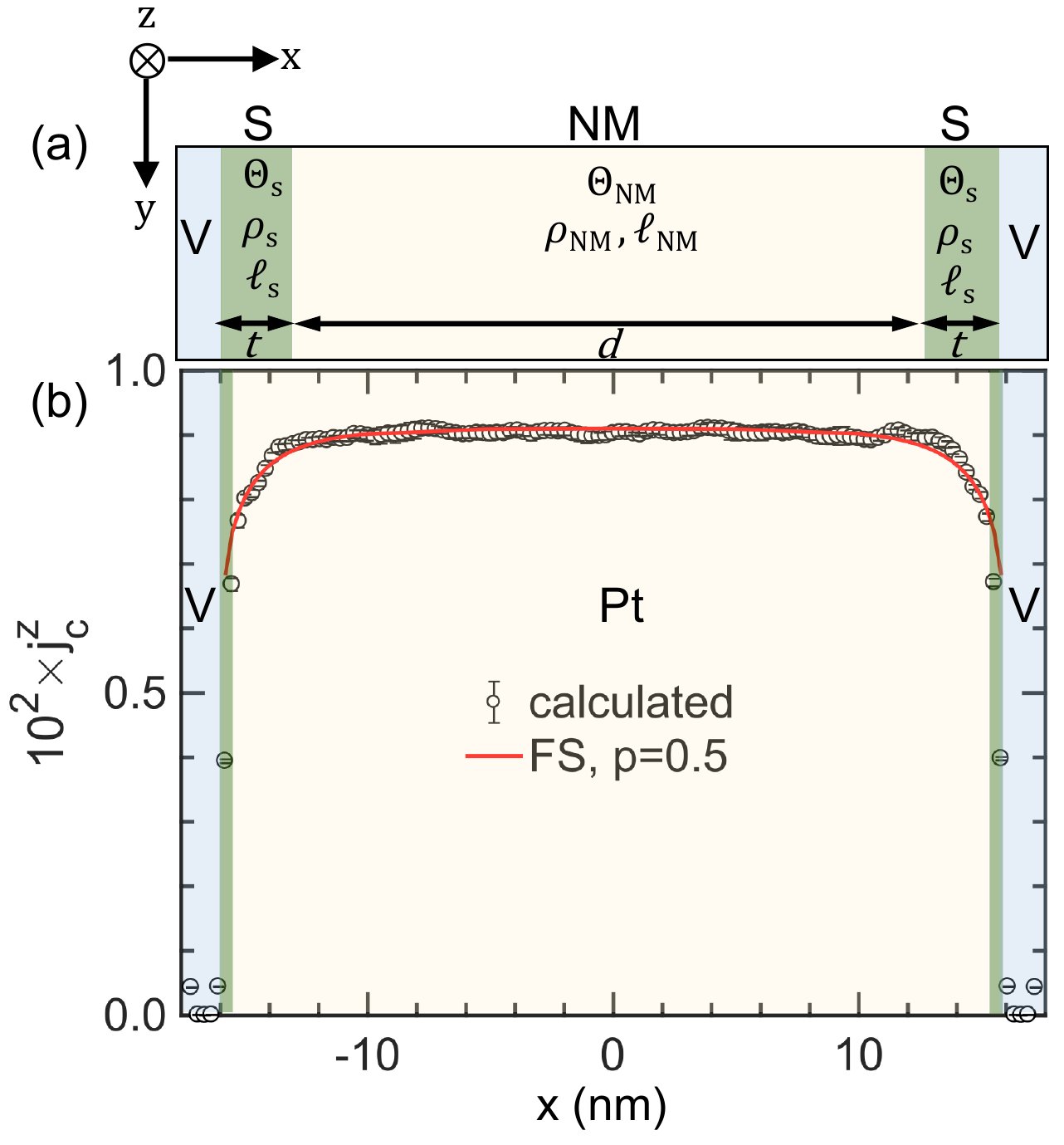}
%\llap{\parbox[b]{6.25in}{\large(a)\\\rule{0ex}{2.85in}}}\llap{\parbox[b]{6.25in}{\large(b)\\\rule{0ex}{1.5in}}}
\caption{(a) Schematic of the  S$|$NM$|$S trilayer geometry with a nonmagnetic (NM) layer sandwiched between two surface layers (S) of finite thickness $t$ that separate it from the vacuum (V). Spin transport in each layer $i$ is characterized by its bulk properties $\Theta_i$, $l_i$ and $\rho_i$. For the surface layer, these transform into surface quantities in the limit $t\rightarrow0$. (b) Profile of the charge current $j_c^z(x)$ flowing through Pt in the $z$ direction averaged in the $y$ direction and sufficiently far from the leads in the $z$ direction as to be independent of $z$. The red curve is a fit using Fuchs-Sondheimer \cite{Fuchs:pcps38, Sondheimer:ap52} theory (FS) in the explicit form given in Ref.~\cite{Lucas:jap65}. Error bars represent the average deviation over five configurations of disorder.}
\label{fig3}
\end{figure}

%%%%%%%%%10%%%%%%%%20%%%%%%%%30%%%%%%%%40%%%%%%%%50%%%%%%%%60%%%%%%%%70%%%%%%%%80 
{\color{red}\it Improved Valet-Fert model.---}To understand the discrepancy, we generalize the Valet-Fert (VF) model of Zhang \cite{Zhang:prl00} to include surface spin-flipping and a surface SHE.
To include surface effects, the thin film geometry is modelled as an S$|$NM$|$S trilayer with surface (S) regions of finite thickness $t$ sandwiching the non-magnetic ``bulk'' metal (NM) of thickness $d$ in the $x$ direction as shown schematically in \cref{fig3}(a). For each layer $i(=$S, NM), the transverse spin current density and spin accumulation excited by a longitudinal charge current density $j^z_c(x)$ in the $z$ direction are
\begin{subequations}
\label{eq3}
\begin{align}
j^x_{sy}(x) &= \frac{1}{2e\rho_il_i}
\Big[ A_ie^{-x/l_i}-B_ie^{x/l_i} \Big] + \Theta_i \, j^z_c(x)    \label{eq3a} \\
\mu^x_{sy}(x) &= A_ie^{-x/l_i}+B_ie^{x/l_i}                      \label{eq3b}
\end{align}
\end{subequations}
where $l_i\equiv l^i_{\rm sf}$ describes the decay of a spin current in the layer $i$, $\rho_i$ is its resistivity and $\Theta_i \equiv \Theta^i_{\rm sH}$ its spin Hall angle. The metallic layer has inversion symmetry and the spin current and spin accumulation are continuous across the $\rm NM|S$ interface. The surface layer is anisotropic and all three transport properties are considered to be tensors with the independent elements $\rho_{\rm S}^{\perp},\rho_{\rm S}^{\parallel},l_{\rm S}^{\perp},l_{\rm S}^{\parallel},\Theta_{\rm S}^{\perp},\Theta_{\rm S}^{\parallel}$. Since we are interested in spin diffusion perpendicular to the surface, $l_{\rm S}^{\parallel}$ and $\Theta_{\rm S}^{\parallel}$ do not enter \eqref{eq3} while $\rho_{\rm S}^{\parallel}$ is implicit in $j^z_c(x)$. We are left with three parameters $\rho_{\rm S}^{\perp},l_{\rm S}^{\perp}$ and $\Theta_{\rm S}^{\perp}$ governing transport perpendicular to the surface in the finite surface layer $\rm S$ that cannot be determined by fitting equations \eqref{eq3} for $j^x_{sy}$ and $\mu^x_{sy}$. Instead, taking the limit $t\rightarrow0$ leads to the surface areal resistance $AR_{\rm S}$ and spin-memory loss (SML) parameter $\delta$ defined as
\begin{align}
\lim_{t\rightarrow0}\rho_{\rm S}^{\perp} t=AR_{\rm S} 
\hspace{1em} \text{and} \hspace{1em} 
\lim_{t\rightarrow0}t/l_{\rm S}^{\perp}=\delta
\end{align} 
where $\delta$ describes the spin current discontinuity that arises when $t\rightarrow0$ and the prefactor of the first term on the right-hand side of \eqref{eq3a} becomes $\delta/2eAR_{\rm S}$. Since there is no transport across the surface into the vacuum, we set $AR_{\rm S} \rightarrow \infty$ so the first term in \eqref{eq3a} vanishes giving rise to the boundary condition $j^x_{sy}(\pm \tfrac{d}{2})=\Theta_{\rm S}^{\perp}j^z_c(\pm \tfrac{d}{2})$ compared to $j^x_{sy}(\pm \tfrac{d}{2})=0$ in \eqref{eq1a}. We use this condition and inversion symmetry in the $x$ direction to eliminate the coefficients $A_i $ and $B_i$ for the slab.

Introducing a surface region whose transport properties differ from those of the bulk poses the question as to how the longitudinal charge transport is affected by the surface. In \cref{fig3}(b), $j^z_c(x)$ is seen to deviate in the topmost $\sim$10 atomic layers from a constant bulk value and can be fit essentially perfectly using the Fuchs-Sondheimer (FS) theory \cite{Fuchs:pcps38, Sondheimer:ap52} with the RT bulk resistivity $\rho_{\rm Pt}$, the mean-free-path $\lambda_{\rm Pt}=3.74\,$nm estimated in the relaxation time approximation with the Fermi velocity averaged over the Fermi surface \cite{Nair:prl21}, and a specularity coefficient $p \sim 0.5$, without needing to explicitly invoke the surface transport parameter $\rho_{\rm S}^{\parallel}$. Using this $j^z_c(x)$ as the longitudinal charge current driving the spin-Hall current in \eqref{eq3a}, we arrive at the following expressions describing spin diffusion in the presence of the spin Hall effect in a thin film
\begin{subequations}
\label{eq4}
\begin{align}
j^x_{sy}(x) &= -C_{\rm S} \cosh(x/l_{\rm Pt}) + \Theta_{\rm Pt} \, j^z_c(x) \label{eq4a}\\
\mu^x_{sy}(x) &= 2e\rho \, l_{\rm Pt}C_{\rm S} \sinh(x/l_{\rm Pt})          \label{eq4b}
\end{align}
\end{subequations}
where the coefficient 
\begin{align}
C_{\rm S}&=
\frac{(\Theta_{\rm Pt}-\Theta^\perp_{\rm S}) j^z_c(|d/2|)}
{\cosh(d/2l_{\rm Pt})}
\label{eq5}
\end{align}
is clearly a surface dependent quantity and 
\begin{equation}
\label{eq6}
\Theta^\perp_{\rm S}= \frac{j^x_{sy}(|d/2|)}{ j^z_c(|d/2|)} . 
\end{equation}
For the [110] surface we find a value of $\Theta^\perp_{\rm S}=2.0\%$, substantially lower than the bulk value. This result as well as those for [001] and [111] oriented films are given in \cref{tab:parameters}. For the very smooth [111] surface we also considered a ``rough'' variant prepared by randomly removing half of the atoms from the topmost layer. The results obtained from \eqref{eq6} are seen to exhibit a very strong dependence on the surface orientation. This comes about not because of a strong dependence of $j^x_{sy}(x)$ on the surface but because of (i) the FS suppression of $j^z_c(x)$ close to the surface and (ii) the connectivity of the atoms making up the surface layer that affects $j^z_c(\pm d/2)$ greatly. For example, $j^z_c(|d/2|)$ has a large value for the smooth [111] surface where every surface atom has six neighbouring surface atoms. It is this which leads to a relatively large value of $j^z_c(|d/2|)$ and low value of $\Theta^\perp_{\rm S}$.

\begin{table}[t]
\caption{Room temperature transport parameters for differently oriented Pt thin films: specularity coefficient $p$; spin Hall angle in percent for spin-current perpendicular ($\Theta_{\rm S}^{\perp}$) and parallel ($\Theta_{\rm S}^{\parallel}$) to the surface. Values of $\Theta_{\rm S}^{\perp}$ were obtained by fitting $j^x_{sy}(x)$ with \eqref{eq4a} and \eqref{eq5} while keeping $\Theta_{\rm Pt}$ and $l_{\rm Pt}$ fixed at their bulk values. For the rough [111] calculations, 50\% of the atoms in the surface layers were removed at random.}
\begin{ruledtabular}
\begin{tabular}{lccccc}
                              &                & \multicolumn{3}{c}{Clean} 
                                                                    & \multicolumn{1}{c}{Rough} \\
\cline{3-5}
                              &                & \multicolumn{1}{c}{[110]} 
                                                      & \multicolumn{1}{c}{[100]} 
                                                             & \multicolumn{1}{c}{[111]} 
                                                                    & \multicolumn{1}{c}{[111]} \\ 
\cline{1-6}	
  $p$                         &                & 0.5 & 0.5 & 0.4 & 0.2 \\
$\Theta_{\rm S}^{\perp}$ (\%) & Eq.\eqref{eq6} & 2.0 & 1.7  & 0.1  & 3.1 \\
        & Eqs.\eqref{eq4a}\&\eqref{eq5}  & 2.1  & 3.8  & 3.5  & 2.0 \\
$\Theta_{\rm S}^{\parallel}$ (\%) &            & 8.8  & 7.0  & 7.7 & 6.3 \\
\end{tabular}
\end{ruledtabular}
\label{tab:parameters}
\end{table}

Just as the resistivity is not a local property on the length scale of the mean free path, there is no reason to describe the SHE as a local property as is done by the semiclassical VF description. At the centre of the thin film furthest from the surfaces, the absolute value of the leading, diffusive term in \eqref{eq4a}  has its minimum value and $j^x_{sy}(x=0)/j^z_c(x=0) \approx \Theta_{\rm Pt}$ as indicated by the dashed blue reference line in \cref{fig2} corresponding to the bulk value of $\Theta_{\rm Pt}=4.2\%$ \cite{Wesselink:prb19}. Fixing $\Theta_{\rm Pt}$ at this bulk value, we scan the $\{l_{\rm Pt}, \Theta^\perp_{\rm S}\}$ parameter space with \eqref{eq4a} and find a minimum occurs for values of $l_{\rm Pt}$ in the range 4--6~nm and of $\Theta^\perp_{\rm S}$ in the range 1.8--2.2\%, \cref{fig2} (bottom right).
Alternatively, we can fix $\Theta_{\rm Pt}$ and $l_{\rm Pt}$ at their bulk values $\{4.2\%,5.2\,$nm\} and optimize the fit of \eqref{eq4a} and \eqref{eq5} to the ab-initio values of $j^x_{sy}(x)$ using the remaining free parameter $\Theta^\perp_{\rm S}$. The result of doing so for the [110] surface is the red curve in \cref{fig2} for a value of $\Theta^\perp_{\rm S}=2.10\pm0.05\%$. The fit is seen to be excellent.
Values of $\Theta^\perp_{\rm S}$ obtained for the other surfaces in this way are given in \cref{tab:parameters} where a suppression of the surface SHA is seen for all orientations.

\begin{figure}[t]
\includegraphics[width=8.6cm]{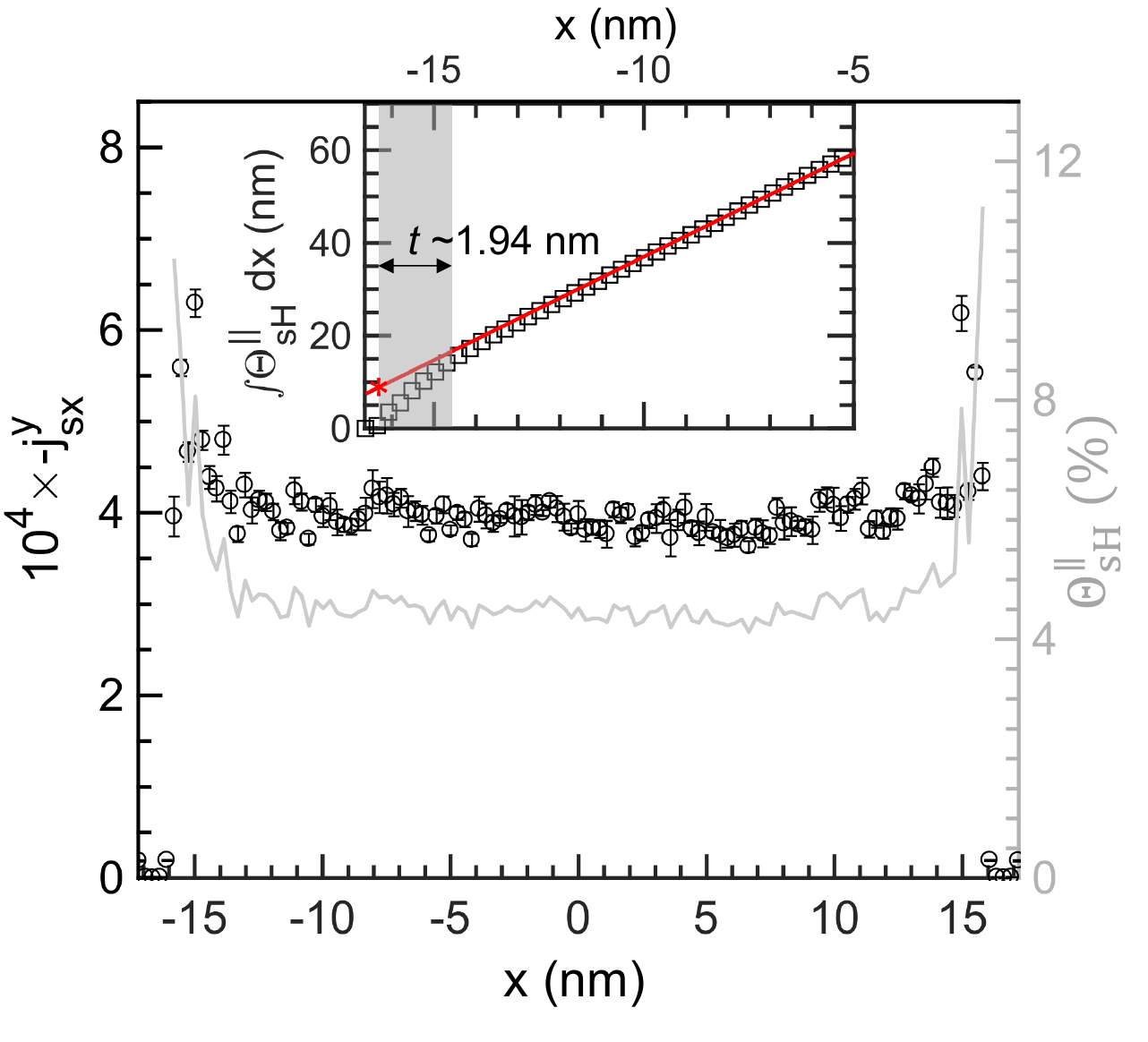}
\caption{RT spin current $-j^y_{sx}$ propagating in the $y$ direction as a function of $x$ (left-hand axis). Error bars represent the average deviation across 5 configurations of disorder. The local ratio $-j^y_{sx}(x)/j^z_c(x) \equiv \Theta_{\rm sH}^\parallel(x)$ plotted as a function of $x$ is referenced to the right-hand axis. Inset: Integral of $\Theta_{\rm sH}^\parallel(x)$ from the centre of the vacuum region near the bottom surface to $x = -5\,$nm, plotted as a function of $x$. The linear fit is shown in red and  the ``effective surface region" in grey. The value of the linear fit at the $x$ coordinate where $\int \Theta_{\rm sH}^\parallel(x) dx$ becomes positive is indicated by a red asterisk.
}
\label{fig4}
\end{figure}

%%%%%%%%%10%%%%%%%%20%%%%%%%%30%%%%%%%%40%%%%%%%%50%%%%%%%%60%%%%%%%%70%%%%%%%%80 
{\color{red}\it Surface enhancement of $j_{sx}^y$.---}We have seen that $j^z_c$ and $j^x_{sy}$ depend strongly on the distance $x$ to a surface in a thin film. In \cref{fig4} we plot the spin current that flows parallel to the surface, $j^y_{sx}$, and see that it also depends quite strongly on $x$. Unlike $j_{sy}^x(x)$ that changes on a length scale of $l_{\rm sf}$, $j^y_{sx}$ is essentially constant up to a mean-free-path of the surface where it increases to reach $\sim 150\%$ of its mean value at the centre of the slab in spite of the FS suppression of the driving charge current $j_c^z(x)$ shown in \cref{fig3}. Thus in a sample that is finite in the $x$ and $y$ directions, see \cref{fig:geometry}, a spin Hall current diffusing towards a surface or an interface will have an additional contribution from the surfaces parallel to the direction of diffusion.  

To extract a surface SHA, we integrate $\Theta_{\rm sH}^\parallel(x)=j^y_{sx}(x)/j^z_c(x) $ from a position in the vacuum where all currents vanish, through the surface at $x=-d/2$, in to $x$ and plot the result as a function of $x$ in the inset to \cref{fig4}. Using a linear fit we identify a surface region of thickness $t \sim 1.94$~nm between where the integrated curve deviates from the linear bulk contribution and where it vanishes. We define the ratio of the corresponding intercept on the $y$ axis (red asterisk in \cref{fig4}) to $t$ as the surface contribution $\Theta^\parallel_{\rm S}$. For the [110] thin film, this ratio yields $\Theta^{\parallel}_{\rm S}\approx 8.8\%$. Results for the other orientations given in \cref{tab:parameters} are all enhanced compared to the bulk value of $\Theta_{\rm Pt}$.

%%%%%%%%%10%%%%%%%%20%%%%%%%%30%%%%%%%%40%%%%%%%%50%%%%%%%%60%%%%%%%%70%%%%%%%%80 
{\color{red}\it Discussion.---}To understand the lateral spin current profile in a thin film, allowance must be made for a variation of the SHA at the surface \cite{WangL:prl16} as well as the Fuchs-Sondheimer suppression of the driving charge current density. When this is done, the spin-current profile that we calculate from first-principles scattering theory is consistent with bulk values of the SHA and SDL that we calculated using the same methodology \cite{Wesselink:prb19, Nair:prl21}. No evidence is found for a recently reported value of $l_{\rm Pt}=11\pm3 \,$nm obtained from MOKE observations of the spin accumulation induced in thin films by the SHE \cite{Stamm:prl17}. The asymptotic ``bulk'' resistivity of 20.6$\, \mu \Omega \,$cm that Stamm {\it et al.} report for thick films \cite{Stamm:prl17} is twice the handbook value quoted for RT Pt \cite{HCP90}. 
Although polycrystallinity and grain boundary scattering are frequently invoked \cite{Mayadas:apl69, Mayadas:prb70} to account for the increased resistivity of thin films \cite{deVries:tsf88, Wu:apl04, Chawla:apl09, Chawla:prb11, Dutta:jap17} and wires \cite{Steinhogl:prb02, Josell:armr09, Chawla:prb11}, it is not known what effect these might have on the SHE-induced spin accumulation nor is any structural characterization reported that might account for the increased resistivity \cite{Stamm:prl17}. From the Elliott-Yafet relation $\rho \, l_{\rm sf}=\,$const., one might expect $l_{\rm sf}$ to be shorter when the resistivity is high rather than longer.  

%%%%%%%%%10%%%%%%%%20%%%%%%%%30%%%%%%%%40%%%%%%%%50%%%%%%%%60%%%%%%%%70%%%%%%%%80
{\color{red}{\it Summary.---}}First-principles calculations are used to examine the diffusion of the intrinsic spin Hall current perpendicular to the surface of a thin film of Pt at room temperature when a charge current is passed through the thin film. 
A marked suppression of the charge current parallel to the surface is found to be well reproduced by a phenomenological model due to Fuchs and Sondheimer and leads to a reduction of the spin-current source. 
A model of spin diffusion that is widely used to interpret experimental SHE studies fails to reproduce the spin current profiles that we calculate leading us to develop a more general model that provides an accurate description of the calculated spin current.
A substantial enhancement of the spin Hall current parallel to the surface is predicted.

%%%%%%%%%10%%%%%%%%20%%%%%%%%30%%%%%%%%40%%%%%%%%50%%%%%%%%60%%%%%%%%70%%%%%%%%80
{\color{red}\it {Acknowledgments.---}}This work was financially supported by the ``Nederlandse Organisatie voor Wetenschappelijk Onderzoek'' (NWO) through the research programme of the former ``Stichting voor Fundamenteel Onderzoek der Materie,'' (NWO-I, formerly FOM) and through the use of supercomputer facilities of NWO ``Exacte Wetenschappen'' (Physical Sciences). R.S.N. acknowledges funding from the Shell-NWO/FOM “Computational Sciences for Energy Research” PhD program (CSER project number 15CSER12). 

%-------10--------20--------30--------40--------50--------60--------70--------80
% Don't keep on deleting this ... just comment it out
%\bibliography{/Users/KellyPJ/Documents/RESEARCH/BibFiles/pjk,notes}
%\bibliography{pjk,notes}

\begin{thebibliography}{49}
\bibitem [{foo({\natexlab{b}})}]{footnote5}%
  \BibitemOpen
  \bibinfo {note} {{With these parameters,
  we obtain bulk values of $l_{\rm Pt}=5.28\pm0.09\,$nm and $\Theta_{\rm
  Pt}=4.27\pm0.03\%$; see Table IV of reference \cite{Wesselink:prb19}. If
  instead of the vBH exchange correlation potential we use the PBE generalised
  gradient approximation \cite{Perdew:prl96}, we obtain values of $l_{\rm
  Pt}=5.23\,$nm and $\Theta_{\rm Pt}=4.54\%$.}}\BibitemShut {Stop}%
\end{thebibliography}

%apsrev4-2.bst 2019-01-14 (MD) hand-edited version of apsrev4-1.bst
%Control: key (0)
%Control: author (8) initials jnrlst
%Control: editor formatted (1) identically to author
%Control: production of article title (0) allowed
%Control: page (0) single
%Control: year (1) truncated
%Control: production of eprint (0) enabled
%

\end{document}